\documentclass[fleqn,12pt,twoside]{article}
\usepackage[headings]{espcrc1}

\readRCS
$Id: espcrc1.tex,v 1.2 2004/02/24 11:22:11 spepping Exp $
\usepackage{graphicx}
\usepackage[figuresright]{rotating}

\newcommand{\AmS}{{\protect\the\textfont2
  A\kern-.1667em\lower.5ex\hbox{M}\kern-.125emS}}

\title{Complex Plasmas as a Model for the Quark-Gluon-Plasma Liquid}

\author{M. H. Thoma\address[MCSD]{Max-Planck-Institute for Extraterrestrial
Physics, \\ 
        P.O. Box 1312, 85741 Garching, Germany}%
        \thanks{Email address: thoma@mpe.mpg.de}}

\begin{document}

% typeset front matter
\maketitle

\begin{abstract}

The quark-gluon plasma, possibly created in ultrarelativistic heavy-ion
collisions, is a strongly interacting many-body parton system. By
comparison with strongly coupled electromagnetic plasmas (classical and
non-relativistic) it is concluded that the quark-gluon plasma could be in
the liquid phase. As an example for a strongly coupled plasma, complex
plasmas, which show liquid and even solid phases, are discussed briefly.
Furthermore, methods based on correlation functions for confirming and
investigating the quark-gluon-plasma liquid are presented. Finally,
consequences of the strong coupling, in particular a cross section
enhancement in accordance with experimental observations at RHIC, are
discussed.

\end{abstract}

\section{Strongly coupled plasmas}

Plasmas are ionized gases forming 99\% of the visible matter in the universe.
Plasmas can be produced by high temperatures, e.g., in stars, by electric
fields, e.g., in neon tubes used for illumination, or by radiation, e.g., in
interstellar plasmas. Plasmas can be classified according to the following
properties: 

1. A plasma can be non-relativistic or relativistic. In the latter case 
the velocity of the plasma particles is close to the speed of light.
Examples are the electron-positron pair plasma in a supernovae or the
quark-gluon-plasma (QGP) in ultrarelativistic heavy-ion collisions.

2. A plasma can be classical or quantum mechanical. In the latter case
the de Broglie wave length of the particles becomes of the order of 
the interparticle distance or larger, which is the case, for instance,
for the QGP. In particular, in degenerate plasmas, such as in white dwarfs,
quantum effects are dominant.

3. A plasma can be ideal or strongly coupled. In the latter case the 
interaction energy between the particles becomes of the order of the kinetic
particle energy or larger. As we will show this is the case for the QGP. 
Another interesting example are complex plasmas, which we will discuss in the
following.

Most plasmas in nature and in the laboratory are non-relativistic,
classical, weakly coupled (ideal) plasmas. As a matter of fact,
strongly coupled plasmas are hard to produce, since they require low 
temperatures and high densities, at which a strong recombination sets in.

For non-relativistic, classical electromagnetic plasmas the Coulomb coupling 
parameter
\begin{equation}
\Gamma = \frac{Q^2}{dT},
\label{eq1}
\end{equation}
where $Q$ is the plasma particle charge, $d$ the interparticle distance, 
and $T$ the plasma temperature, distinguishes between weakly and 
strongly coupled plasmas. In the latter case $\Gamma \geq O(1)$ holds.
Most plasmas are ideal with $\Gamma <10 ^{-3}$. Examples for strongly
coupled plasmas are the ion component in white dwarfs \cite{ref1} or 
short-living
high-density plasmas produced with heavy-ion beams on solid state targets
at GSI \cite{ref2}.

Obviously, strongly coupled plasmas require a non-perturbative description, 
e.g., molecular dynamics in the classical case. This has been used for
computing the equation of state of 
one-component plasmas with a pure (repulsive) Coulomb interaction. It turned
out that for $\Gamma > 172$ a Coulomb crystal forms \cite{ref3}, 
i.e., it is energetically 
more favorable if the particles arrange in a regular structure.

\bigskip

\includegraphics[width=9cm]{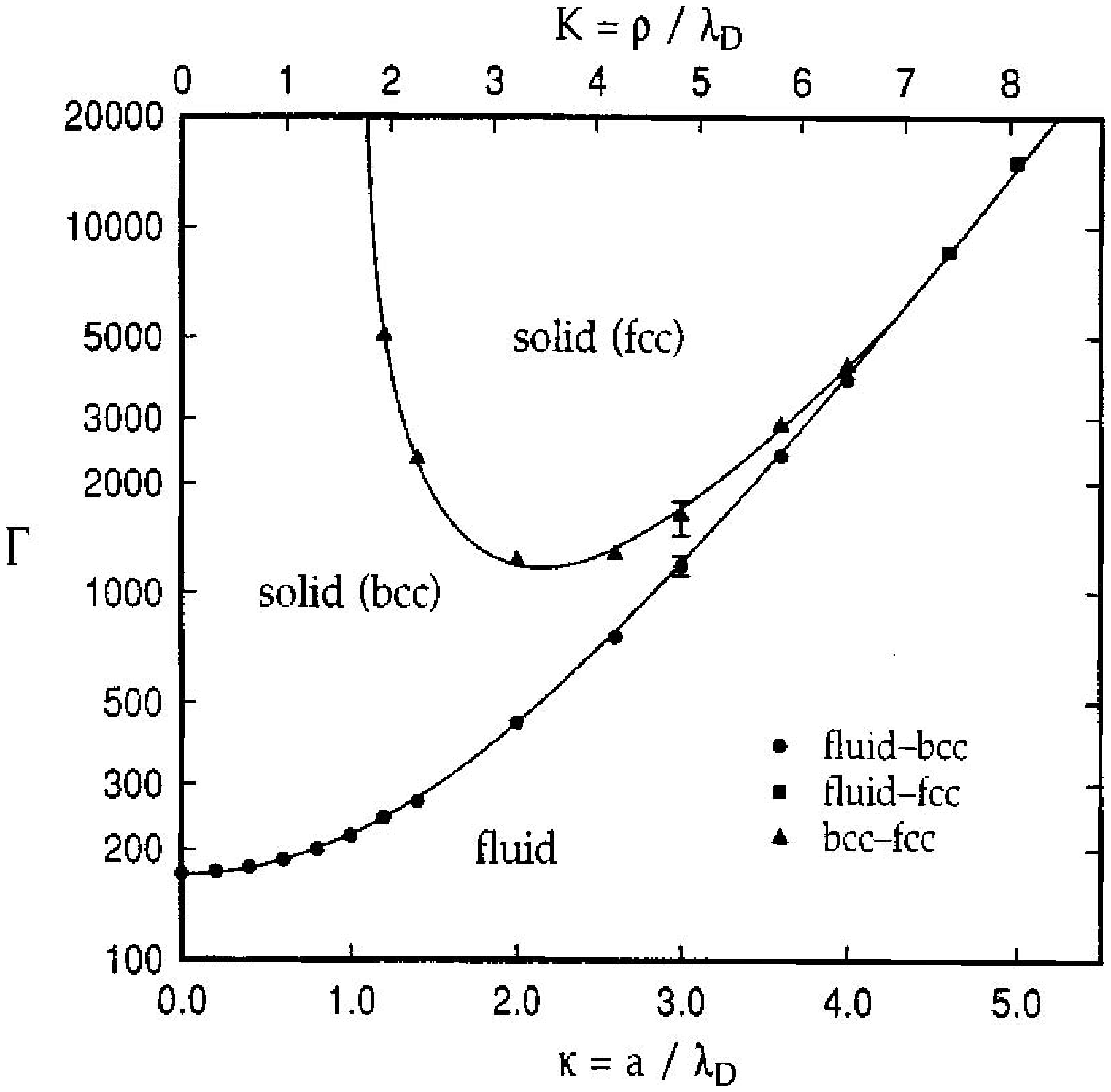}

Fig.1: Phase diagram of a strongly coupled Yukawa system from Ref.\cite{ref4}.

\bigskip

Under most circumstances, however, Debye screening cannot be neglected and
plasmas are Yukawa systems with an interaction potential
\begin{equation}
V(r)=\frac{Q}{r}\> e^{-r/\lambda_D}
\label{eq2}
\end{equation}
with the Debye screening length $\lambda_D$. Then it is convenient to 
introduce a second parameter $\kappa = d/\lambda_D$. In Fig.1 the phase diagram
in the $\Gamma$ - $\kappa$ plane for a repulsive, one-component 
Yukawa system is shown \cite{ref4}. One recognizes a solid state, 
either bcc or fcc,
for large $\Gamma$ and small $\kappa$. Otherwise a liquid phase exists.
Note, however, that in the case of a purely repulsive interaction there is
no gas-liquid transition but only a supercritical fluid, i.e., a clear 
distinction between the gaseous and the liquid state is not possible. 
However, in the strongly coupled case, $\Gamma > O(1)$, the system behaves
more like liquid than a gas, as we will discuss below. 

\section{Complex Plasmas}

Dusty or complex plasmas are multi component plasmas containing ions,
electrons, neutral gas, and microparticles, e.g., dust grains. Such 
a situation can be studied, for example, in a low temperature noble
gas discharge plasma, in which micron size particles, e.g., monodisperse
plastic spheres with a diameter of 1 - 10 $\mu$m, are injected.
Due to the high mobility of the electrons, the microparticles collect 
mostly electrons on their surface, leading to a large negative
charge of the microparticles between 10$^3$ and 10$^5$
$e$ depending on the grain size and electron temperature ($T_e= 1 -10$ eV). 
Since the interparticle distance is typically of the order of 200 $\mu$m
and the kinetic energy of the particles corresponds to room temperature
due to friction with the neutral gas,
$\Gamma \gg 1$ (up to values of 10$^5$) can easily be reached.
Therefore, in 1986 it was predicted that a regular ordering in
the charged microparticle system, regarded as a massive plasma component, 
could exist in complex laboratory plasmas \cite{ref5}. This prediction
triggered experimental efforts to search for this new state of matter,
the so called plasma crystal, which
was discovered for the first time in 1994 at the Max-Planck-Institute for
Extraterrestrial Physics \cite{ref6} and almost at the same time at 
other places \cite{ref7,ref8}. 

An image (top view) of such a plasma crystal is shown in Fig.2. 
Here a plasma chamber with an extension of a few centimeters is used,
in which a low-temperature argon plasma is ignited by a rf discharge.
After injecting the microparticles a plasma crystal is formed within 
a few seconds. For observation a single
horizontal layer is illuminated by a laser sheet and the scattered light 
(Mie scattering) is recorded by a CCD camera. Clearly a hexagonal structure
with lattice defects similar as in a real crystal can be observed. 

\bigskip

\includegraphics[width=10cm]{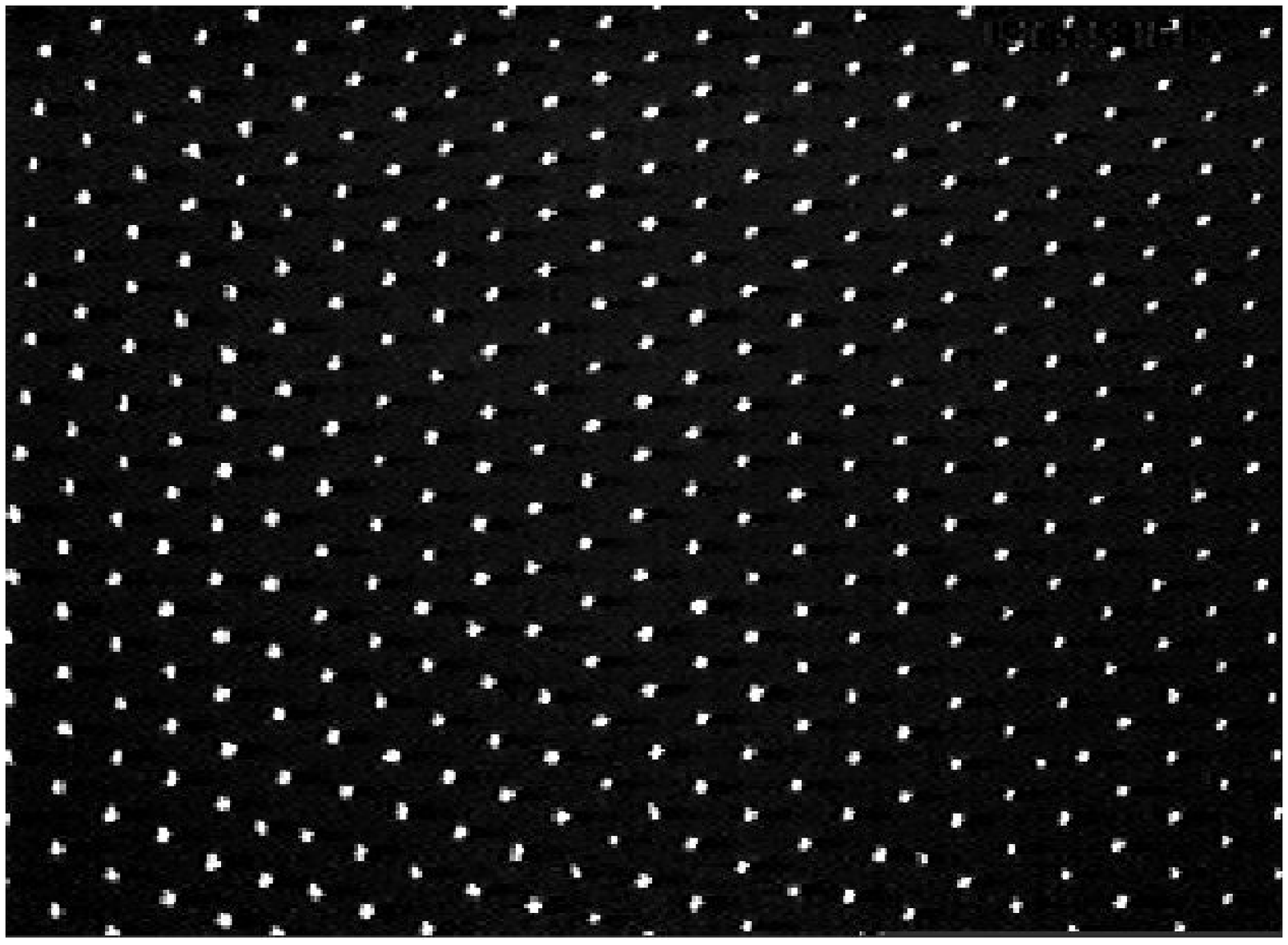}

Fig.2: Top view of the plasma crystal.

\bigskip

Phase transitions to the liquid phase and the disordered gas phase can 
be observed by reducing the pressure \cite{ref9}. This leads to a 
reduction of the
neutral gas friction and hence an increase of the temperature of the 
microparticle system. At the same time the electron temperature decreases
leading to a reduction of the charge of the microparticles. 
Hence the Coulomb coupling parameter (\ref{eq1}) decreases and can become 
smaller than its critical value for crystallization. Indeed 
reducing, for instance, the pressure from about 60 Pa, 
in which a crystal structure exists, to 30 Pa, the microparticles show
a liquid behavior. Going to even lower pressure, e.g., 10 Pa, the
velocity of the particles is increased by a factor of about 200
compared to the solid phase, in which the thermal lattice oscillations
correspond to particle velocities of the order of 0.2 mm/s. Thus at 10 Pa 
the system resembles a gas.

The equation of state of a complex plasma and the value of the 
Coulomb coupling parameter can be determined by the so called
pair correlation function, defined as 
\begin{equation}
g({\bf r}) = \frac{1}{N}\> \langle \sum_{i\neq j}^N \delta 
({\bf r}+{\bf r_i}-{\bf r_j}) \rangle,
\label{eq3}
\end{equation}
where $N$ is the particle number and ${\bf r_i}$ and ${\bf r_j}$
the positions of the particles.    

In the case of a regular crystal structure, where a long range order
exists, the pair correlation function shows pronounced peaks at the
locations corresponding to the nearest neighbors, next to nearest
neighbors and so on. Of course, due to thermal fluctuations and defects
the peaks have a finite width and their height decreases with increasing
distance. 
In the case of a liquid, where only a short range order corresponding
to a fixed interparticle distance in the incompressible fluid is 
present, the pair correlation function exhibits only one clear peak and 
some times one or two small and broad additional peaks.  
Finally, in a gas corresponding to a disordered system no peaks show up
in the pair correlation function.

Before we turn to the QGP, let me mention that the investigation of complex
plasmas is hampered by the presence of gravity in the laboratory. The
microparticles can be suspended against gravity only by an electric field
(typically a few V/cm) in a small region of the plasma chamber, the
plasma sheath above the bottom where the field can be strong enough. 
Hence only small systems in vertical
directions, e.g., quasi 2-dimensional crystals, can be built up. 
Furthermore, the plasma 
conditions in the plasma sheath are very complicated rendering the 
interpretation of the results difficult. Finally, the gravitational force 
is comparable to the interparticle force, thus strongly disturbing
the system, e.g., the crystal structure, and making some measurements even
impossible. Therefore we perform experiments with complex plasmas also under
microgravity conditions, i.e., in parabolic flights, in sounding rockets, 
and on board of the International Space Station where we installed the
first scientific experiment (PKE-Nefedov) \cite{ref10}.

Finally, let me discuss some applications of complex plasmas. Beside 
investigating the plasma crystal and its melting a large variety of different 
experiments can de conducted, e.g., phonons and plasma waves can be
excited, instabilities can be observed, shock waves and Mach cones can be
produced, shear flow can be studied, etc. In general, complex plasmas
are ideal model systems to study the dynamical behavior of solids,
liquids, and plasmas on the microscopic level in real time. In addition,
charged dust systems and dust-plasma interactions play an important role
in astrophysics (comets, interstellar clouds, planet formation, etc.),
in environmental research (chemical reactions involving
micrometeorites in the ionosphere), as well as in plasma technology 
(dust contamination in the micro-chip production using plasma etching,
dust in fusion reactors, etc.). 

\section{Applications to the quark-gluon plasma}

For applying the ideas and methods used for strongly coupled
electromagnetic plasmas, such as complex plasmas, to the 
physics of the QGP, we first estimate the interaction parameter. 
In the case of QCD it reads \cite{ref11}
\begin{equation}
\Gamma \simeq 2 \frac{C\alpha_s}{dT},
\label{eq4}
\end{equation}
where $C=4/3$ or $C=3$ is the Casimir invariant of quarks or gluons,
respectively. The factor 2 in (\ref{eq4})
comes from the fact that in ultrarelativistic systems the magnetic
interaction is as important as the electric.  
Assuming a temperature $T=200$ MeV which corresponds to
a coupling constant $\alpha_s = 0.3$ to 0.5 and an interparton distance
$d$ of about 0.5 fm, we find $\Gamma = 1.5$ - 6. Hence the QGP is
a strongly coupled plasma at temperatures which can be reached
in accelerator experiments, and it is conceivable that it behaves more 
like a liquid than a gas. Indeed RHIC data (elliptic flow, particle spectra)
\cite{ref11a}, 
which can be described
well by hydrodynamics with a negligible viscosity and which suggest
a fast thermalization, indicate such a behavior 
\cite{ref12,ref13,ref14,ref15}. 

Therefore
the phase diagram of hot hadronic matter could be possibly extended
by another phase transition from the liquid to the gas phase
of the QGP at a temperature of maybe a few hundred MeV as sketched 
in Fig.3.
However, this requires an attractive as well as repulsive component
of the potential, e.g., a Lennard-Jones type potential, which is not
the case in QCD where the interaction between partons in the various
channels is either attractive or repulsive \cite{ref16}. 
However, it is known that in a strongly coupled plasma purely repulsive 
forces can obtain an attractive component at certain distances due
to non-linear effects \cite{ref17}. Hence it might be worthwhile to look
for a gas-liquid transition in heavy-ion collisions, in particular at LHC.
Regardless whether this phase transition exists or whether there is only
a supercritical QGP liquid, the QGP should behave more like a liquid
at temperatures close to the deconfinement transition and like a gas at
temperatures far above this transition.

\bigskip

\includegraphics[width=10cm]{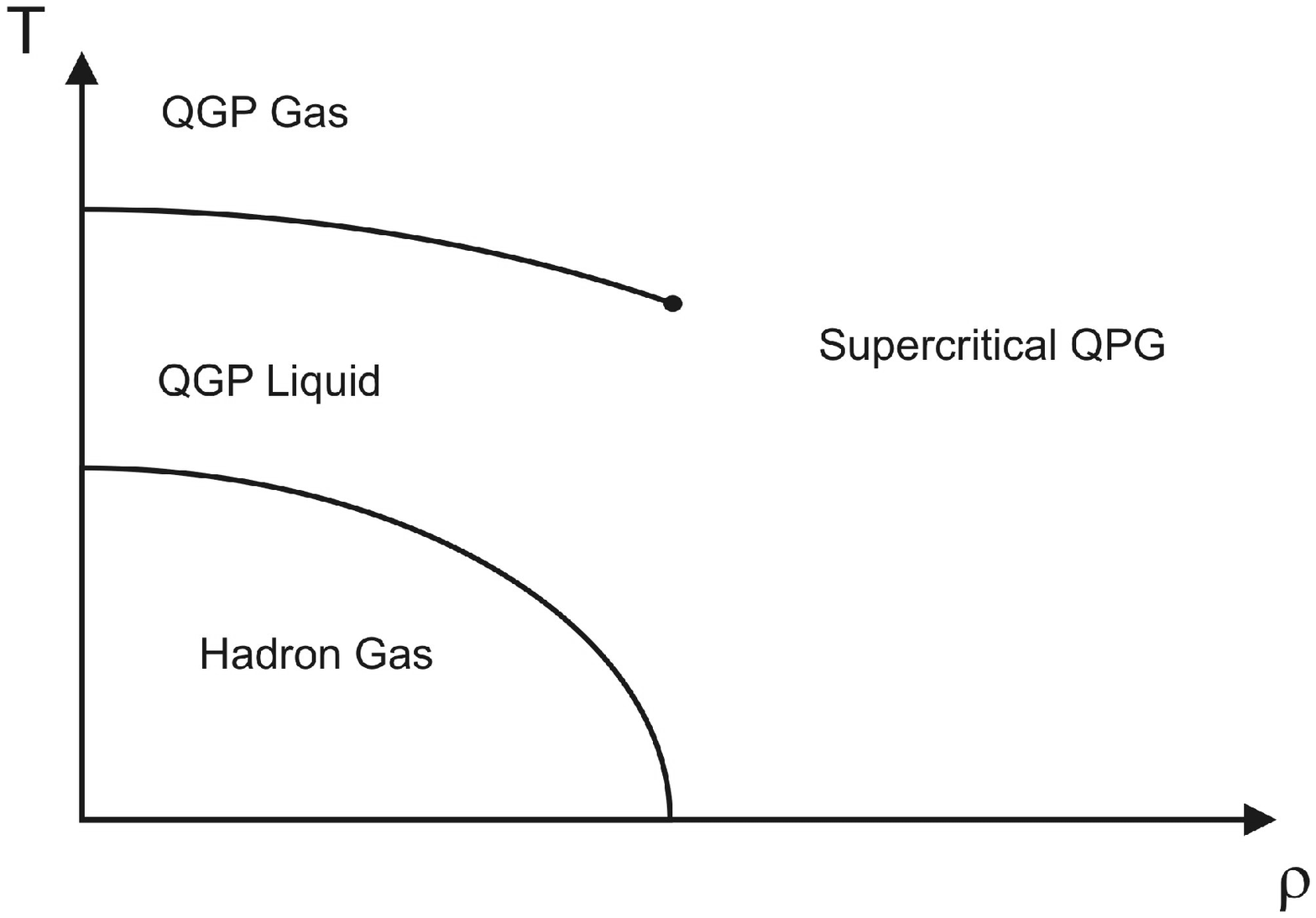}

Fig.3: Sketch of a phase diagram of hadronic matter with a possible
gas-liquid transition in the QGP phase.

\bigskip

For verifying and investigating the liquid state 
quantitatively one could consider
experimentally as well as theoretically the so called static structure
function which is closely related to the Fourier transform of the pair
correlation function (\ref{eq3}). The static structure function is a standard tool
for the experimental and theoretical analysis of liquids \cite{ref18}.
The qualitative behavior of the static structure functions for
liquids and gases is shown in Fig.4. In the case of a liquid the static 
structure function shows oscillations with decreasing amplitudes for
large momenta. For an interacting gas, on the other hand, the static 
structure function increases monotonically reaching quickly a saturation 
value.

\bigskip

\includegraphics[width=10cm]{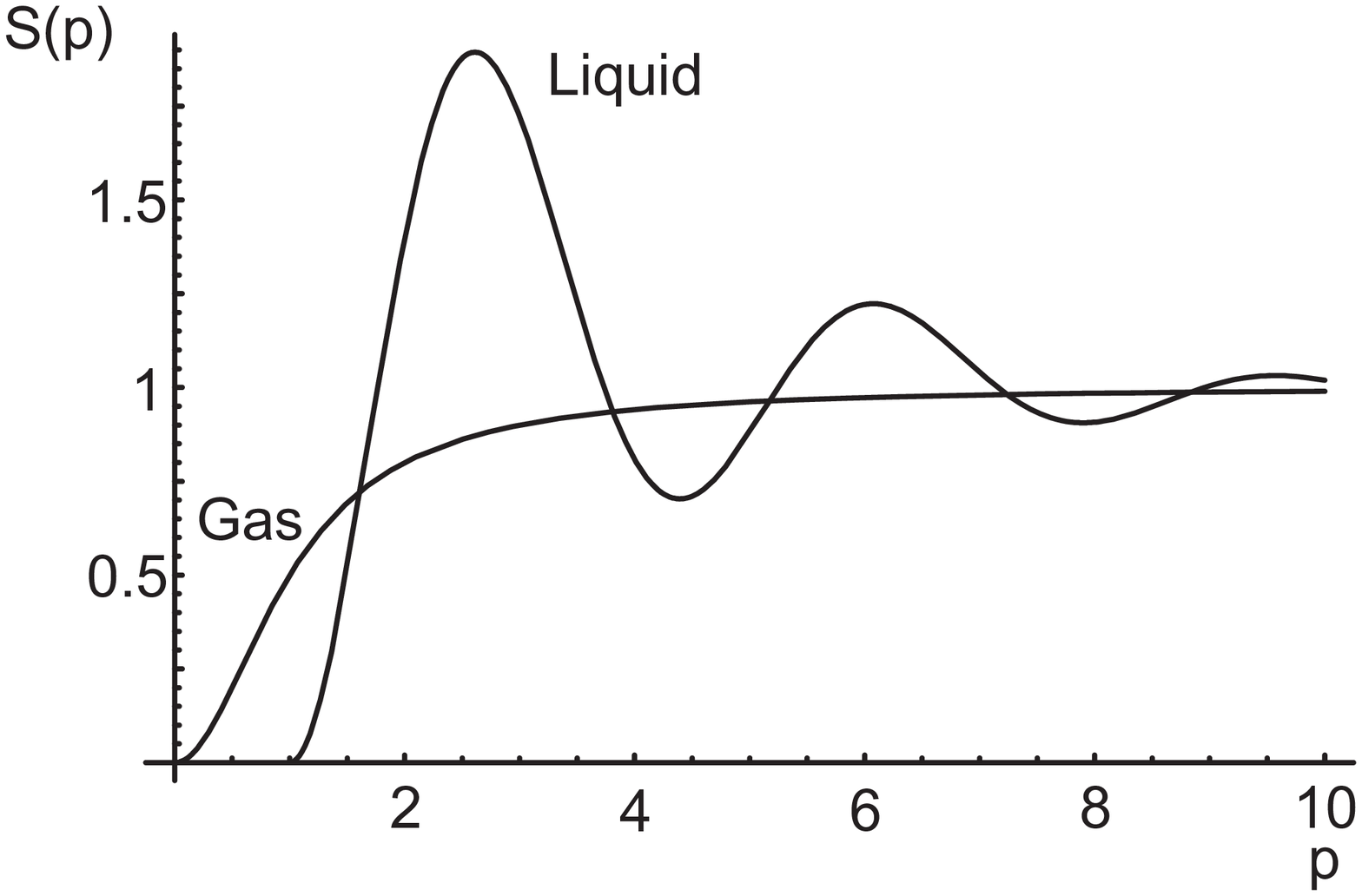}

Fig.4: Qualitative behavior of the static structure functions for a liquid
and a gas.

\bigskip

In Ref.\cite{ref19} we defined the static structure function
for the case of the QGP and showed that it is related to the longitudinal
part of the QCD polarization tensor. Furthermore we
argued that QCD lattice simulations should
be able to prove the liquid behavior of the strongly coupled QGP
by computing the static structure function. 
To demonstrate the use of this definition and as 
a reference for lattice calculations, we have calculated the
static structure function within the Hard Thermal Loop (HTL) approximation,
yielding \cite{ref19}
 \begin{equation}
S(p) 
=\frac{2N_fT^3}{n}\> \frac{p^2}{p^2+m_D^2},
\label{eq5}
\end{equation}
where $N_f$ is the number of light quark flavors, $n$ the parton density, and
$m_D=1/\lambda_D$ the Debye screening mass which is proportional to
$gT$ in the HTL approximation. This $p$-dependence clearly belongs
to an interacting gas which is not surprising as the HTL approximation
is based on the high-temperature assumption, $T\gg T_c$.

The pair correlation function follows from the Fourier 
transform  of $S(p)-1$ as
\begin{equation}
g(r) = -\frac{N_fT^3}{2\pi n}\> \frac{m_D^2}{r}\> e^{-m_Dr},
\label{eq6}
\end{equation}
showing no peaks which corresponds, of course, also to the gas phase.

Finally, we want to point out that strongly coupled plasmas show in general
a cross section enhancement for the interaction of the particles 
within the plasma. The reason is that the Coulomb radius, defined by
$r_c=Q^2/E$ with the particle energy $E$, in a strongly coupled plasma
is of the order of the Debye screening length or even larger. Hence
the standard Coulomb scattering theory has to be modified since due to the
strong interaction particles outside of the Debye sphere contribute 
significantly and the Debye screening length cannot be used as an infrared
cutoff. This modification leads, for example, to the experimentally observed
enhancement of the so called ion drag force in complex plasmas which is
caused by the ion-microparticle interaction \cite{ref20}.

In the QGP at $T\simeq 200$ MeV, the ratio $r_c/\lambda_D = 1$ - 5,
leading to a parton cross section enhancement in the QGP by a factor
of 2 - 9 \cite{ref11} compared to perturbative results. Additional 
cross section enhancement could come from non-linear (modification of the 
Yukawa potential to a power law potential at large distances) and 
non-perturbative effects.
An enhanced parton cross section leads to a reduced mean free path $\lambda $
of the partons in the QGP which corresponds
to a small viscosity $\eta \sim \lambda$ and a fast thermalization
as indicated by RHIC experiments.

A cross section enhancement of the elastic parton
scattering by an order of magnitude compared to perturbative results 
has also been postulated 
in Ref.\cite{ref20a} by considering elliptic flow data and 
particle spectra observed at RHIC. The assumption of a strongly coupled
QGP, which requires an infrared cutoff smaller than the Debye mass
and in which non-linear and non-perturbative effects are important,
gives a natural explanation for this enhancement.

Another consequence would be the enhancement of the collisional
energy loss 
\begin{equation}
\left (\frac{dE}{dx}\right )_{\rm coll} \simeq \frac{\Delta E}{\lambda },
\label{eq7}
\end{equation}
where $\Delta E$ is the energy transfer per collision. This contribution
to the total energy loss receives an enhancement from the soft part 
due to the mean free path reduction for low-energy partons \cite{ref20b}.
On the other hand,
it can be expected that the radiative energy loss is suppressed in
the strongly coupled QGP by the Landau-Pomeranchuk-Migdal effect.
In the case of a dense and strongly coupled plasma the emission 
of gluons could be suppressed by this formation time effect. 
Indeed experimental indications for explaining jet quenching with a
significant contribution from the collisional energy loss 
have been discussed recently \cite{ref21,ref22}.

\section{Conclusions} 

Within the last years strongly coupled plasmas became of increasing 
interest in fundamental research as well as in technology applications.
The QGP as well as complex plasmas are important examples of strongly
coupled plasmas. Definitely, the QGP is the most challenging strongly coupled 
plasma from the experimental as well as theoretical point of view.
Complex plasmas, on the other hand, can easily be produced and studied.
Therefore they can be used as ideal models to investigate fundamental
aspects of other many-body systems, such as the dynamic evolution of phase 
transitions on the microscopic level. Complex plasmas 
can also be used to learn some interesting lessons about the QGP 
by analogy, such as the existence of a possible liquid phase and
a gas-liquid transition, the use of correlation functions for investigating
the equation of state, or cross section enhancement in a strongly coupled 
plasma.

\end{document}